\documentclass[%
 reprint,
superscriptaddress,
 amsmath,amssymb,
 aps,
 prb,
]{revtex4-2}
\usepackage{graphics}
\usepackage{graphicx}
\usepackage{epsfig}
\usepackage{amssymb}
\usepackage{amsthm}
\usepackage{xcolor}
\usepackage{hyperref}

\begin{document}
\title{Using quantitative magneto-optical imaging to reveal why the ac susceptibility of superconducting films is history-independent}

\author{Davi A. D. Chaves}
\email{davi@df.ufscar.br}
\affiliation{Departamento de F\'{\i}sica, Universidade Federal de S\~{a}o Carlos, 13565-905 S\~{a}o Carlos, SP, Brazil}

\author{J. C. Corsaletti Filho}
\altaffiliation[Present address: ]{Laboratório Nacional de Luz Síncrotron, Centro Nacional de Pesquisa em Energia e Materiais, 13083-100 Campinas, SP, Brazil}
\affiliation{Departamento de F\'{\i}sica, Universidade Federal de S\~{a}o Carlos, 13565-905 S\~{a}o Carlos, SP, Brazil}

\author{E. A. Abbey}
\affiliation{Departamento de F\'{\i}sica, Universidade Federal de S\~{a}o Carlos, 13565-905 S\~{a}o Carlos, SP, Brazil}

\author{D. Bosworth}
\affiliation{Department of Materials Science and Metallurgy, University of Cambridge, Cambridge CB3 0FS, United Kingdom}

\author{Z. H. Barber}
\affiliation{Department of Materials Science and Metallurgy, University of Cambridge, Cambridge CB3 0FS, United Kingdom}

\author{M. G. Blamire}
\affiliation{Department of Materials Science and Metallurgy, University of Cambridge, Cambridge CB3 0FS, United Kingdom}

\author{T. H. Johansen}
\affiliation{Department of Physics, University of Oslo, POB 1048, Blindern, 0316 Oslo, Norway}

\author{A. V. Silhanek}
\affiliation{Experimental Physics of Nanostructured Materials, Q-MAT, CESAM, Université de Liège, B-4000 Sart Tilman, Belgium}

\author{W. A. Ortiz}
\affiliation{Departamento de F\'{\i}sica, Universidade Federal de S\~{a}o Carlos, 13565-905 S\~{a}o Carlos, SP, Brazil}

\author{M. Motta}
\email{m.motta@df.ufscar.br}
\affiliation{Departamento de F\'{\i}sica, Universidade Federal de S\~{a}o Carlos, 13565-905 S\~{a}o Carlos, SP, Brazil}

\date{\today}
\begin{abstract}

Measurements of the temperature-dependent ac magnetic susceptibility of superconducting films reveal reversible responses, i.e., irrespective of the magnetic and thermal history of the sample. This experimental fact is observed even in the presence of stochastic and certainly irreversible magnetic flux avalanches which, in principle, should randomly affect the results. In this work, we explain such an apparent contradiction by exploring the spatial resolution of magneto-optical imaging. To achieve this, we successfully compare standard frequency-independent first harmonic ac magnetic susceptibility results for a superconducting thin film with those obtained by ac-emulating magneto-optical imaging (acMOI). A quantitative analysis also provides information regarding flux avalanches, reveals the presence of a vortex-antivortex annihilation zone in the region in which a smooth flux front interacts with pre-established avalanches, and demonstrates that the major impact on the flux distribution within the superconductor happens during the first ac cycle. Our results establish acMOI as a reliable approach for studying frequency-independent ac field effects in superconducting thin films while capturing local aspects of flux dynamics, otherwise inaccessible via global magnetometry techniques.

\end{abstract}

\maketitle

\section{Introduction}

The last years have seen superconducting materials be positioned as a vital part of an ongoing quantum revolution~\cite{MacFarlane2003,Devoret2013,Wendin2017,Kjaergaard2020,Blais2021} and serving as a fertile playground for the development of several nanoscale technological applications~\cite{Natarajan2012,McCaughan2014,Zhao2018,Strambini2020,Chen2020,Ligato2022,Golod2022,Chaves2023}. Particularly, understanding, controlling, and exploring the interaction of different superconductors with distinct properties and structures with a low-frequency ac magnetic field has been an active research topic~\cite{Clem1994,Willemin1998,Brandt2002,Hoffmann2011,Geng2015,Silva2016,Ivan2018,Shen2020,Pasquini2021}.

In a type-II superconductor, it may be energetically favorable for the sample to allow flux penetration in the form of vortices~\cite{Abrikosov1957,Blatter1994}. For a given direction of the applied magnetic field, vortices may either be of positive or negative polarity, the latter being commonly referred to as antivortices. Whereas vortices with the same polarity interact repulsively~\cite{Brandt2009}, vortices and antivortices attract each other, which eventually leads to mutual annihilation when two such entities come in close proximity~\cite{Chaves2011,Sardella2009}.

On a mesoscopic scale, ordinary flux distributions in type-II specimens are described by critical state models~\cite{Bean1964,Kim1962}. In this case, the magnetic field gradually penetrates toward the center of the sample as a smooth flux front originating from the edges of the material, a consequence of vortex motion being hampered by pinning centers~\cite{Jooss2002,Colauto2020}. The exact distribution profile depends on sample geometry and its magnetic history~\cite{Zeldov1994,Brandt1998,Shantsev1999}. Moreover, the depth of the flux front penetration is tied to the sample critical current density $J_c$, as further penetration indicates a lower magnetic shielding capability~\cite{Zeldov1994}. In short, the actual flux distribution usually depends on external thermodynamic parameters such as the temperature, $T$, and the applied magnetic field, $H$, i.e., $J_c = J_c(T,H)$.

The inevitable vortex displacement during flux penetration represents an energy dissipating process~\cite{Blatter1994}. Then, if the superconductor is not able to swiftly assimilate the heat generated by moving vortices in order to accommodate for further vortex movement, a thermomagnetic instability may be triggered. In a given interval of magnetic fields and temperatures, these events lead to the onset of a positive feedback process in which superconducting properties are locally suppressed, allowing for abrupt flux penetration known as flux avalanches~\cite{Wipf1991,Denisov2006a,Denisov2006b}. In thin films, flux avalanches take remarkable dendritic patterns as they propagate through the material with velocities up to the scale of hundreds of km/s~\cite{Leiderer1993,Duran1995,Johansen2001,Bolz2003,Aranson2005,Vestgarden2011,Colauto2015,Vestgarden2018}.

The abrupt flux penetration during a flux avalanche event results in well-known flux jumps in the global magnetization hysteresis loop of superconductors~\cite{Schawlow1956,Kim1963,Esquinazi1999,Altshuler2004,Colauto2020}. Another signature of avalanches in the magnetic properties of superconducting materials is a paramagnetic reentrance observed in the temperature dependence of the first harmonic ac magnetic susceptibility, $\chi_{\text{ac}}(T) = \chi_{\text{ac}}'(T) + i\chi_{\text{ac}}''(T)$~\cite{Silhanek2004,Menghini2005,Motta2011}. The in-phase component $\chi_{\text{ac}}'$ is related to an inductive response and measures the superconductor ability to shield magnetic flux~\cite{Goldfarb1991,Gomory1997}. The so-called paramagnetic reentrance is observed as a decrease in $|\chi_{\text{ac}}'|$ for temperatures lower than the superconducting critical temperature $T_c$. On its turn, the out-of-phase component $\chi_{\text{ac}}''$ gauges the energy losses related to flux motion in type-II superconductors~\cite{Goldfarb1991,Gomory1997}. Hence, an increase in  $\chi_{\text{ac}}''$ accompanying the decrease in $|\chi_{\text{ac}}'|$ reveals the occurrence of flux avalanches. Although ac susceptibility studies are a ubiquitous approach for characterizing the magnetic dynamics of superconducting systems~\cite{Muller1989,Goldfarb1991,Gomory1997,Oliveira2010,Kramer2010,Zhao2012,Silva2016,Topping2019,eremets2022high,Ghigo2022}, a technique with the micrometric spatial resolution of magneto-optical imaging, has remained little explored in this effort.

In this work, we investigate the effects of ac magnetic fields in a 100-nm-thick amorphous MoSi (a-MoSi) film by employing ac-emulating magneto-optical imaging (acMOI). Comparing the results with $\chi_{\text{ac}}(T)$ measurements obtained by conventional global ac magnetometry, we demonstrate that acMOI is a reliable technique for the quantitative study of the ac magnetic susceptibility of superconductors. Moreover, as magneto-optical imaging allows to spatially resolve individual flux avalanches, acMOI is used to explain an observed thermomagnetic history-independent paramagnetic reentrance in $\chi_{\text{ac}}(T)$ for the a-MoSi film. Quantitative acMOI also allows us to visualize how an incoming smooth flux front overrides the flux distribution of pre-established avalanches, revealing a vortex-antivortex annihilation zone separating regions permeated by magnetic flux with opposing polarities.

This paper is organized as follows: Section~\ref{Sc:experimental} details the experimental methods used to fabricate and investigate the a-MoSi thin film; Section~\ref{Sc:acSusc-MPMS} describes typical $\chi_{\text{ac}}(T)$ measurements conducted in a standard ac magnetometer, demonstrating the history-independent paramagnetic reentrance in the investigated sample; Section~\ref{Sc:acSusc-MOI} qualitatively explores the nature of ac susceptibility measurements using acMOI both in the smooth penetration and avalanche regimes and quantifies the magnetic imprint of individual avalanches; Section~\ref{Sc:qMOI} demonstrates how acMOI may be used to quantitatively gauge $\chi_{\text{ac}}(T)$ for superconducting samples; Section~\ref{Sc:ErasingAval} further explores the spatial resolution of MOI to investigate how an incoming flux front interacts with an already established avalanche region; finally, Section~\ref{Sc:conclusions} summarizes the results and outlines perspectives on the use of acMOI.

\section{Experimental details}
\label{Sc:experimental}

A square a-MoSi film with lateral size of 2.5~mm and thickness of 100~nm was deposited onto a silicon substrate at 77~K by dc magnetron sputtering at a pressure of 1.2~Pa in dynamical equilibrium under argon flow, similarly to the protocol described in Ref.~\cite{Bosworth2015}. Amorphous MoSi films typically present critical temperatures above 7~K, low intrinsic pinning, and correspondingly low critical current densities~\cite{Kubo1988,Banerjee2017}. Application-wise, a-MoSi is a prominent material choice for superconducting nanowire single-photon detectors~\cite{Verma2015,Caloz2018,Zhang2021}.

The complex ac magnetic susceptibility of the a-MoSi sample was investigated as a function of temperature using standard global magnetometry which captures the magnetic behavior of the sample as a whole. A SQUID-based magnetometer model MPMS-5S from Quantum Design (MPMS) was employed to measure both the in-phase ($\chi_{\text{ac}}'$) and out-of-phase ($\chi_{\text{ac}}''$) components of $\chi_{\text{ac}}(T)$. Probe ac magnetic fields of frequencies $f$ = 0.05~Hz or 1~Hz and amplitudes $h$ varying from 0.1~Oe to 3.8~Oe were applied perpendicularly to the plane of the film during the experiments. Before measuring, the magnetic history of the dc field-generating superconducting coil of the MPMS was erased, and all measurements were performed under remanent magnetic field, $H_{\text{rem}} \lesssim$ 1~Oe. In other words, no external dc field was intentionally applied to the sample.

The magneto-optical imaging technique allows us to locally resolve the magnetic flux distribution within the sample on the micrometric scale~\cite{Jooss2002}. By placing a Bi-doped yttrium iron garnet\textemdash a Faraday-active material~\cite{Helseth2001}\textemdash directly on top of our superconducting film, MOI allows inspection of the deviation of the polarization angle of light in the presence of a magnetic field due to the Faraday effect. Thus, we are able to detect subtle nuances in the local field induced in the investigated material as a variation in the intensity captured by a CCD camera.

We perform a pixel-by-pixel calibration procedure implemented on MATLAB to obtain quantitative information from magneto-optical images~\cite{Shaw2018}. In other words, we extract the out-of-plane magnetic flux density $B(x,y)$ from the intensity data $I(x,y)$, where $(x,y)$ defines the position of a given pixel within the image. Possible drifts in sample position relative to the sensors due to thermal dilation of the cold finger in the experimental setup are corrected within a precision of two pixels using the StackReg plugin~\cite{Thevenaz1998} with ImageJ~\cite{Schneider2012}. As a consequence of its lower $J_c$, a-MoSi also presents an intrinsic advantage for quantitative MO studies, as it inhibits unwanted magnetic domain wall switching in the garnet layer~\cite{Ferrari2007}, which could otherwise compromise the $I$-to-$B$ transformation.

\section{Ac susceptibility: MPMS measurements}
\label{Sc:acSusc-MPMS}

\begin{figure}[htbp]
\centering
\includegraphics[width=\columnwidth,keepaspectratio]{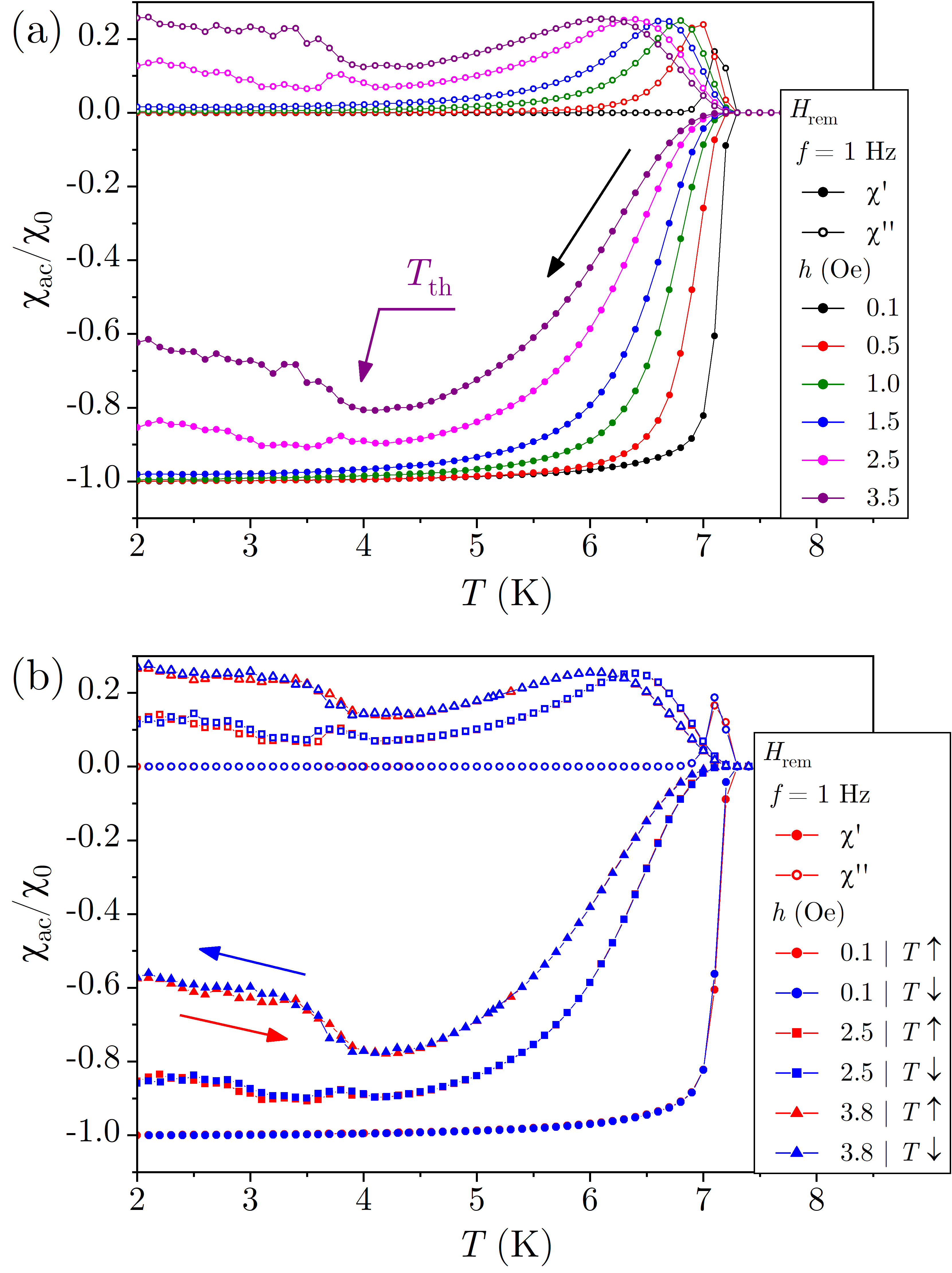}
\caption{Temperature-dependent ac susceptibility of a-MoSi film under $H_{\text{rem}}$ obtained using the MPMS. (a) Data acquired as the temperature is decreased from the normal state using a probe magnetic field of $f$ = 1~Hz and amplitude varying from $h$ = 0.1~Oe to $h$ = 3.5~Oe. Purple arrow indicates the onset of flux avalanches, characterizing the paramagnetic reentrance region. (b) Data acquired both as the temperature is decreased from the normal state ($T \downarrow$) and increased from the Meissner state ($T \uparrow$) with $f$ = 1~Hz and $h$ = 0.1~Oe (smooth regime) and $h$ = 2.5~Oe and 3.8~Oe (avalanche regime).}
\label{Fig:ACSusc}
\end{figure}

Typical temperature-dependent ac susceptibility results for superconducting films are illustrated in Fig.~\ref{Fig:ACSusc} for the 100-nm-thick a-MoSi sample. The curves depicted are obtained using the MPMS and show both $\chi_{\text{ac}}'$ and $\chi_{\text{ac}}''$ normalized by the Meissner state plateau $\chi_0$ of the $\chi_{\text{ac}}'$ measurement conducted with the lowest $h$ and $T$. Figure~\ref{Fig:ACSusc}(a) highlights the effects of $h$ in $\chi_{\text{ac}}(T)$. In all measurements, the sample is first subjected to zero-field-cooling (ZFC) down to 2~K. Then, $\chi_{\text{ac}}(T)$ is measured as the temperature is increased using a probe field with $f$ = 1~Hz and varying $h$ values from 0.1~Oe to 3.5~Oe. As we demonstrate in Appendix~\ref{App:frequency}, the choice of $f$ has almost no impact on the $\chi_{\text{ac}}(T)$ behavior in the frequency range explored in this work. For the smallest field amplitude (black points), we observe a near constant $\chi_{\text{ac}}'$ close to $-1$ at low temperatures. This is a signature of superconductors' perfect diamagnetism, showing that the sample initially shields its interior from magnetic flux very efficiently. Then, as the sample approaches its critical temperature ($T_c$), a sharp increase in $|\chi_{\text{ac}}'|$ is observed, as the film is no longer shielded from flux penetration. Signatures of the superconducting-normal state transition are also found in the out-of-phase component, as a peak in $\chi_{\text{ac}}''$ accompanies the increase in $|\chi_{\text{ac}}'|$. Thus, the dissipative motion of vortices entering the sample is consistently captured by $\chi_{\text{ac}}''$, which is greatly enhanced during flux penetration. Therefore, for the a-MoSi sample, we define $T_c = 7.30 \pm 0.05$~K as the first experimental point for which both $\chi_{\text{ac}}'$ and $\chi_{\text{ac}}''$ depart from zero in $\chi_{\text{ac}}(T)$ measurements.

If $h$ is now increased to 0.5~Oe (red points), a very similar behavior is observed in Fig.~\ref{Fig:ACSusc}(a). However, flux exclusion becomes less complete as the temperature and field are increased in accordance with critical state models. Therefore, although $T_c$ is unchanged, the onset of the superconducting-normal state transition occurs for lower temperatures, as $T$ is increased from 2~K. This trend continues as $h$ is further increased to 1.0~Oe and 1.5~Oe, represented in Fig.~\ref{Fig:ACSusc}(a) by green and blue points, respectively. For all measurements with $h \leq$ 1.5~Oe, the a-MoSi film is in the smooth penetration regime and all flux penetration occurs gradually and uniformly from the edges toward the center of the sample, as described by critical state models. For $h$ = 2.5~Oe, however, a radically different behavior is observed: the pink points in Fig.\ref{Fig:ACSusc}(a) sharply differ from those observed in the smooth regime. For the lowest temperatures, an apparently noisy response is observed in both $\chi_{\text{ac}}(T)$ components while shielding becomes much less effective. The purple points reveal the same trend for measurements carried out with a probe field of 3.5~Oe. As we will demonstrate in this paper using MOI, these characteristics are signs of the occurrence of magnetic flux avalanches in the film~\cite{Raedts2004}. Such variations in $|\chi_{\text{ac}}'|$ and $\chi_{\text{ac}}''$ are then explained by a reduction of the volume of the film free from flux penetration as avalanches advance throughout the sample. Eventually, as $T$ is increased above 4~K, the noisy behavior in $\chi_{\text{ac}}(T)$ is no longer present for both the 2.5~Oe and 3.5~Oe curves. This occurs because the temperature is increased beyond that for which avalanches can be triggered ($T_{\text{th}}$), keeping the sample in a thermomagnetically stable condition, as described by the thermomagnetic model~\cite{Yurchenko2007}. As such, flux will now only penetrate the sample smoothly, although frozen imprints of previous avalanches may remain in the flux patterns observed in the film.

Flux avalanches are of a stochastic nature. It is therefore not possible to accurately predict their shape or size, nor to precisely pinpoint the moment or the position at which an avalanche will be triggered~\cite{Denisov2006b,Colauto2015}. In spite of this fact, Fig.~\ref{Fig:ACSusc}(b) reveals an interesting feature of $\chi_{\text{ac}}(T)$ measurements: the results are not only largely reproducible but also independent of the thermomagnetic history both in the smooth and in the unpredictable avalanche regime. To illustrate that, we conduct $\chi_{\text{ac}}(T)$ measurements for the a-MoSi film after ZFC to 2~K using a probe field with $f$ = 1~Hz and $h$ = 0.1~Oe. The red circles represent the results as $T$ is gradually increased through $T_c$ up to 9~K. Then, $\chi_{\text{ac}}(T)$ is recorded as the temperature is lowered from the normal state back to 2~K, as shown by the blue circles. A close inspection of both curves reveals essentially no difference in $\chi_{\text{ac}}(T)$ in the smooth regime, independently of the direction of the temperature variation. If now $h$ is increased to 2.5~Oe and the experiment is repeated, the sample is in the avalanche regime for $T < T_{\text{th}} \sim 4.5$~K. In this temperature range, the red and blue squares in Fig.~\ref{Fig:ACSusc}(b) are no longer indistinguishable, although they remain very close to each other. More precisely, the observed ups and downs in $\chi_{\text{ac}}(T)$ gauged as $T$ is increased mirror those obtained as $T$ is decreased. The red and blue triangles in Fig.~\ref{Fig:ACSusc}(b) reveal the same behavior in the avalanche regime for a higher probe field amplitude $h$ = 3.8~Oe.

\section{Ac susceptibility: MOI measurements}
\label{Sc:acSusc-MOI}

We now turn to magneto-optical imaging to explain why there appears to be to a large extent a reversible response in the noisy ac susceptibility behavior caused by stochastic avalanche events. To do so, it is instructive to first recall the working principle of how $\chi_{\text{ac}}(T)$ is obtained in magnetometers such as the MPMS. An applied zero-mean probe ac field with an amplitude $h$ and frequency $f$, such that $h(t) = h\cos(2\pi ft)$, induces a time-dependent magnetic moment in the investigated sample. Hence, a detectable electric current is induced in the magnetometer's superconducting pickup coils, connected to the SQUID sensor, allowing the determination of the magnetic moment $m_{\text{ac}}$. After averaging measurements performed for successive probe field cycles, $m_{\text{ac}}$ is fitted to an equation of the form~\cite{QD-ACoption,*QD-IntroAC}
\begin{equation}
    m_{\text{ac}} = C(t) + m'\cos(2\pi ft) + m''\sin(2\pi ft),
\label{Eq:MPMS-acFit}
\end{equation}
where $C(t)$ represents any dc offset or drift in field or temperature, and $m'$ and $m''$ are respectively related to $\chi_{\text{ac}}'$ and $\chi_{\text{ac}}''$ as
\begin{equation}
   \chi_{\text{ac}}' = \frac{m'}{h} \hspace{12pt} \text{and} \hspace{12pt} \chi_{\text{ac}}'' = \frac{m''}{h}.
\label{Eq:MPMS-m'm''}
\end{equation}
Recalling that $\chi_{\text{ac}} = \chi_{\text{ac}}' + i\chi_{\text{ac}}'' = \partial M/\partial H$, if the total applied magnetic field is $H = H_{\text{dc}} + h$, then $\chi_{\text{ac}} = \partial M/\partial h$. Hence, the above measurement protocol can be used in combination with a dc applied magnetic field to gauge the sample susceptibility in different points of the $M(H)$ curve.

\begin{figure}[htbp]
\centering
\includegraphics[width=\columnwidth,keepaspectratio]{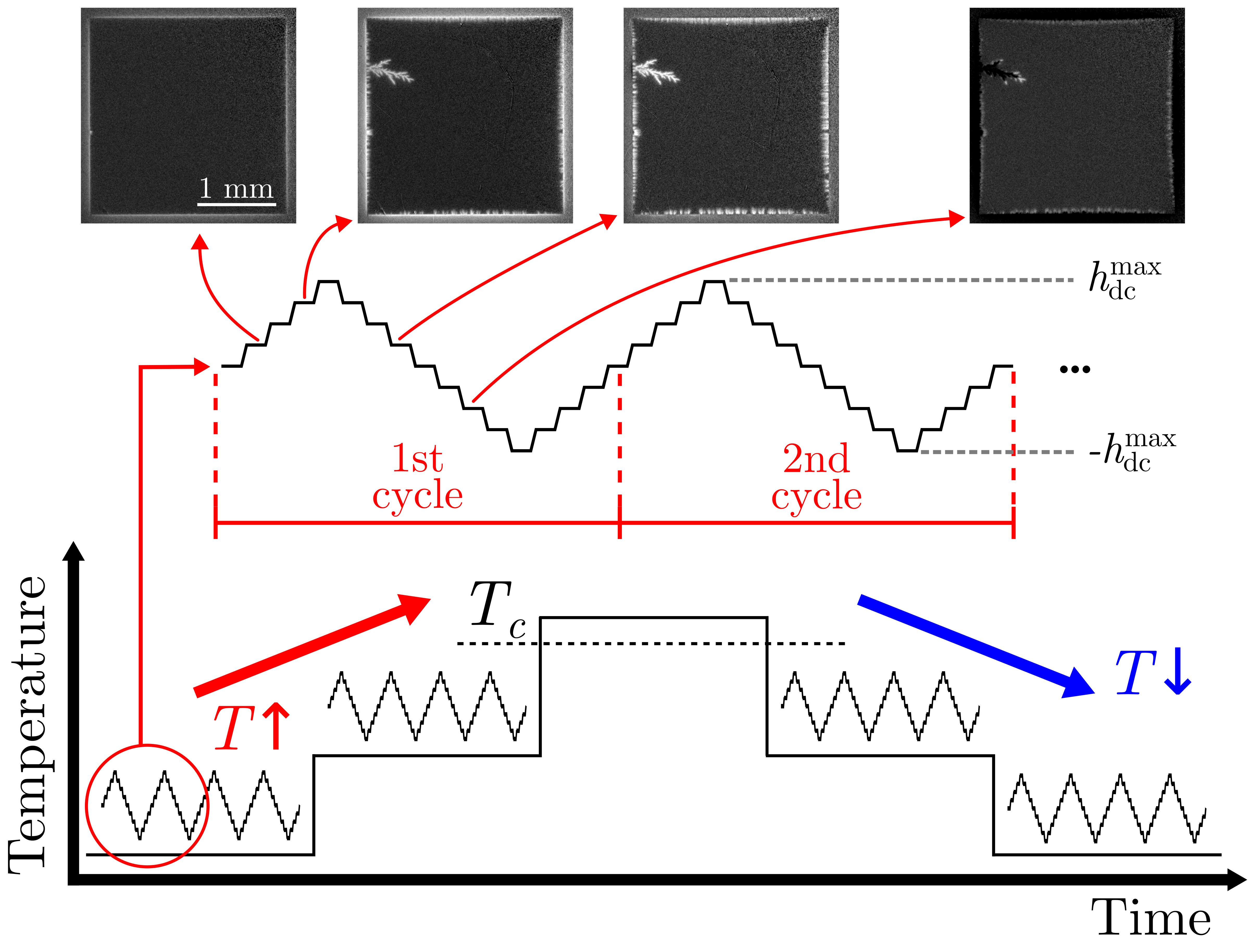}
\caption{Ac-emulated magneto-optical imaging (acMOI). A dc magnetic field is progressively applied in stair-like discrete steps. The dc field intensity is varied from zero to $h_{\text{dc}}^{\text{max}}$, then from $h_{\text{dc}}^{\text{max}}$ to $-h_{\text{dc}}^{\text{max}}$, and finally from $-h_{\text{dc}}^{\text{max}}$ to zero. Successively repeating this field cycle emulates an applied low-frequency ac magnetic field. After each field step, a MO image of the sample is recorded, as exemplified in the detail of the first field cycle. Additional parameters may be controlled, as schematically represented by an increase followed by a decrease in the temperature.}
\label{Fig:acMOI}
\end{figure}

The process to emulate ac measurements using a magneto-optical imaging setup equipped with a dc magnetic field source was first introduced by Ref.~\cite{Motta2011}. Here, we refer to such measurements as acMOI. To summarize the process, the dc field is incremented in stair-like steps until it reaches a preset maximum amplitude $h_{\text{dc}} = h_{\text{dc}}^{\text{max}}$. After each field step, a MO image is recorded. Then, keeping the same step size, the applied field is reduced to $-h_{\text{dc}}^{\text{max}}$ and, finally, increased to zero. This routine reproduces one ac field cycle and it is schematically presented in Fig.~\ref{Fig:acMOI}. Although the data acquisition rate of the acMOI technique is substantially slower than the MPMS ac magnetic field source, by successively repeating the above routine, we may take advantage of the frequency-independent nature of the first harmonic $\chi_{\text{ac}}$ to capture ac effects in the investigated sample. We are also capable of varying external parameters, such as the temperature or an additional dc field, allowing investigations of their effects on the sample. Following, we will explore this ability to qualitatively visualize how magnetic flux penetrates the superconducting a-MoSi film during typical temperature-dependent ac susceptibility measurements, both in the smooth and in the avalanche regimes.

\subsection{Smooth penetration regime}
\label{MOI-smooth}

\begin{figure*}[htbp]
\centering
\includegraphics[width=\textwidth,keepaspectratio]{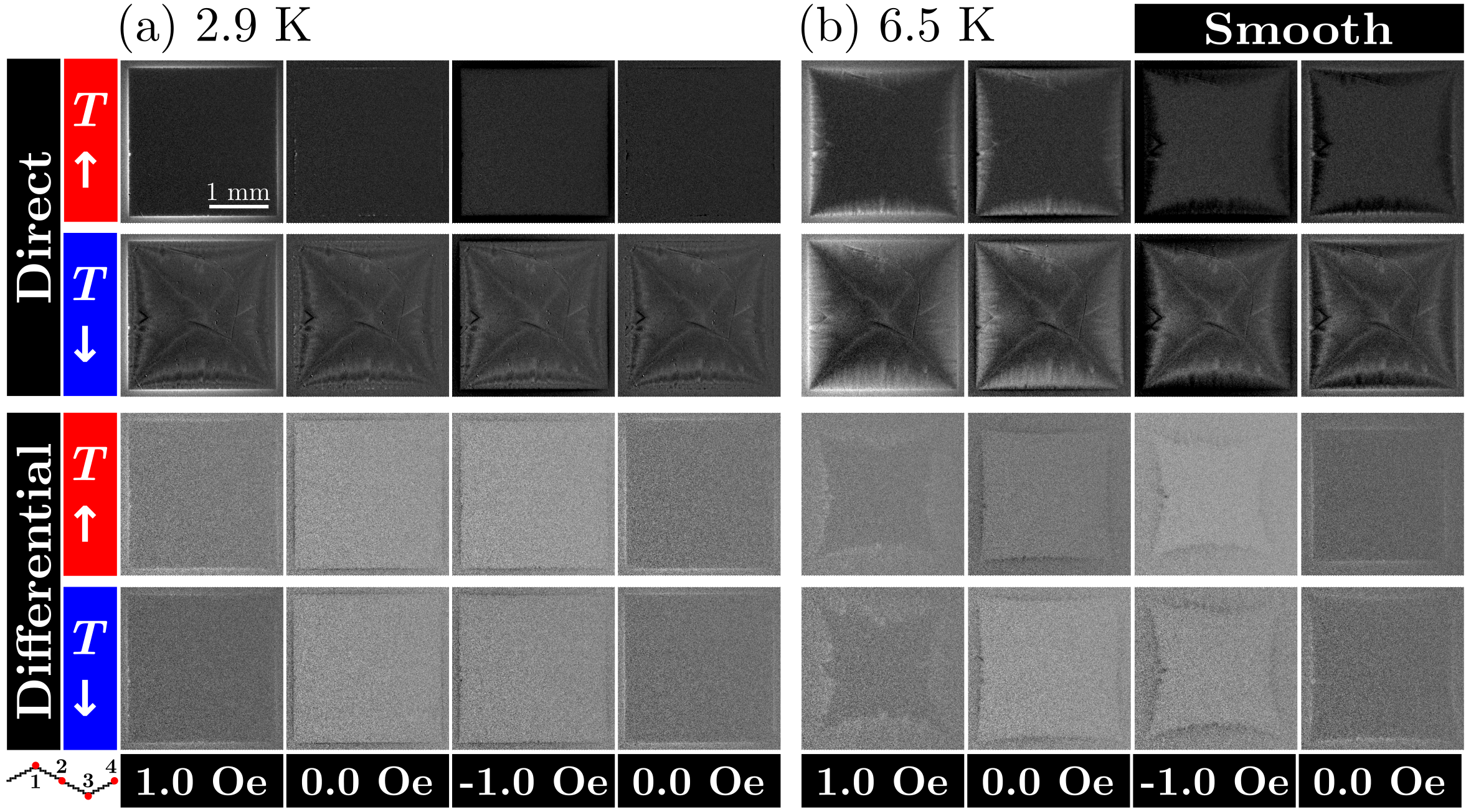}
\caption{Comparison between direct MO images and differential MO images of a-MoSi film taken at (a) 2.9~K and (b) 6.5~K as the temperature is increased ($T \uparrow$) after ZFC to the base temperature and as it is decreased ($T \downarrow$) from above $T_c$. Data is acquired using an ac-emulating applied field with an amplitude of 1.0~Oe, thus in the smooth regime. The contrast in each image was individually adjusted for optimal visualization of the flux penetration. The lower left corner detail represents the point of the field cycle at which each image was captured.}
\label{Fig:Reg-vs-Diff-Smooth}
\end{figure*}

Figure~\ref{Fig:Reg-vs-Diff-Smooth} exemplifies results obtained for the a-MoSi film following the acMOI procedure. In this case, $h_{\text{dc}}^{\text{max}}$ = 1.0~Oe, corresponding to the situation in which the film remains in the smooth regime for all temperatures below $T_c$, as revealed by Fig.~\ref{Fig:ACSusc}(a). The first row of Fig.~\ref{Fig:Reg-vs-Diff-Smooth}(a) shows MO images as directly obtained during the first field cycle after the sample was zero-field-cooled to the base temperature of 2.9~K. A schematic representation of the point in the field cycle at which each of the four images is captured is presented at the lower left corner of Fig.~\ref{Fig:Reg-vs-Diff-Smooth}. The first image reveals a shallow bright region surrounding the darker inner region of the square film at 1.0~Oe. As discussed previously, such a bright region represents the small flux front able to penetrate the superconductor at lower temperatures and ac field amplitudes, due to its elevated shielding capacities. As the field cycle continues, the second image, taken at 0~Oe, reveals that some positive flux remains trapped in the sample, but the edges of the film no longer appear in bright contrast as the flux polarity is being reversed in that region. In the third image, taken at $-1.0$~Oe, the flux inside the superconductor has completely reversed its sign and appears now in dark contrast, signaling its negative intensity. Finally, the fourth image, at 0~Oe, reveals some trapped negative flux in the interior of the sample, but the edges again indicate the reversal of the applied field. The ac-emulated field cycle is repeated a total of four times before the temperature is increased and set to 3.5~K, 4.0~K, 4.5~K, 5.0~K, 5.5~K, 6.0~K, 6.5~K, and 7.0~K, every time repeating the field cycle four times and collecting a MO image after each field step of 0.1~Oe. We will refer to this data as the $T \uparrow$ experiment. The first row of images of Fig.~\ref{Fig:Reg-vs-Diff-Smooth}(b) shows results obtained at 6.5~K as the temperature is increased after ZFC. They are analogous to those obtained at 2.9~K, however, the flux front penetrates deeper into the film due to its reduced shielding capability near $T_c$.

Then, the temperature is risen above $T_c$ in the absence of an applied magnetic field, erasing the magnetic history of the sample. After that, $T$ is progressively reduced back to the base temperature while subjecting the sample to four ac-emulating field cycles at the same set temperatures listed before. This is the $T \downarrow$ experiment. The second row in Fig.~\ref{Fig:Reg-vs-Diff-Smooth}(a) shows the MO images recorded at 2.9~K during this experiment, i.e., after $T$ was reduced from above $T_c$. Although the temperature is the same, the flux landscapes inside the superconductor in the first and second rows of Fig.~\ref{Fig:Reg-vs-Diff-Smooth}(a) are completely different. For the images taken during the $T \downarrow$ experiment, the complete magnetic history of the sample due to the successive field cycles is retained by the film. This happens because higher temperatures enable further flux penetration, therefore the flux trapped in the most inner regions of the sample is not superimposed by new field cycles at lower temperatures. Accordingly, the MO images obtained at 6.5~K as $T$ is reduced, shown in the second row of Fig.~\ref{Fig:Reg-vs-Diff-Smooth}(b), differ from those presented in the first row. In this case, the sample was previously subjected to four ac-emulating field cycles at 7.0~K, resulting in the observed trapped flux in the interior of the film.

After these observations, it may be natural to ask why such different flux distributions lead to the indistinguishable $\chi_{\text{ac}}(T)$ observed in the smooth regime in Fig.~\ref{Fig:ACSusc}(b) for increasing and decreasing temperatures. To understand this, it is necessary to remember that ac susceptibility is a measurement of the flux variation in the material as the applied field is changed, rather than its total magnetic moment. To gauge flux variation due to the variation of an applied field using MOI, we may turn to what is called differential MOI~\cite{soibel2000imaging}. This approach consists of subtracting the measured flux density distribution in a given MO image from that obtained in the previous field step. In other words, $B^{\text{diff}}_{n}(x,y) = B_{n}(x,y) - B_{n-1}(x,y)$, where $n$ represents the MO image number, chronologically increasing from the first to the last image obtained in a given data set.

The third and fourth rows of Figs.~\ref{Fig:Reg-vs-Diff-Smooth}(a) and (b) show differential MO images of the same images represented in the first two rows of each figure. The results demonstrate that, although the flux distribution in the superconductor is vastly different depending on its thermomagnetic history, the flux variation within an ac field cycle does not present significant differences for measurements conducted while increasing or decreasing the sample temperature\textemdash given that the probe field amplitude and frequency are kept the same. This notion explains why $\chi_{\text{ac}}(T)$ does not depend on the sample's thermomagnetic history in the smooth regime. In the Supplemental Material~\cite{[{See Supplemental Material at }][{ for videos highlighting aspects of the flux penetration dynamics revealed in the main text.}]supp}, a video highlights this behavior for all MO images obtained during the four field cycles at temperatures of 2.9~K, 3.8~K, 4.5~K, 5.5~K, and 6.5~K.

\subsection{Flux avalanches regime}
\label{MOI-aval}

\begin{figure*}[htbp]
\centering
\includegraphics[width=\textwidth,keepaspectratio]{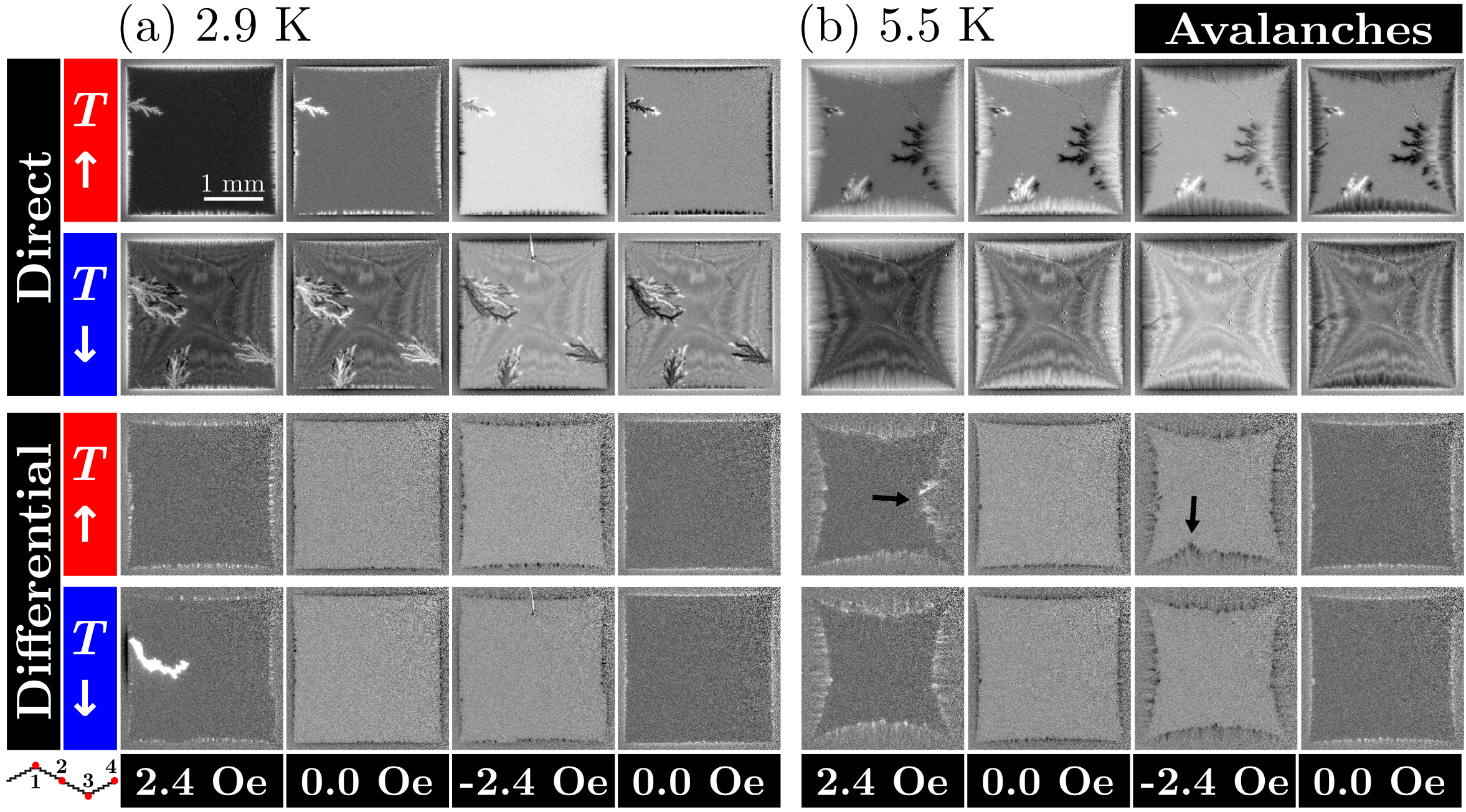}
\caption{Comparison between direct MO images and differential MO images of a-MoSi film taken at (a) 2.9~K and (b) 5.5~K as the temperature is increased ($T \uparrow$) after ZFC to the base temperature and as it is decreased ($T \downarrow$) from above $T_c$. Data is acquired using an ac-emulating applied field with an amplitude of 2.4~Oe, thus in the avalanche regime. The contrast in each image was individually adjusted for optimal visualization of the flux penetration. The lower left corner detail represents the point of the field cycle at which each image was captured. Black arrows indicate regions of further flux penetration that will be discussed in Section~\ref{Sc:ErasingAval}.}
\label{Fig:Reg-vs-Diff-Aval}
\end{figure*}

Figure~\ref{Fig:Reg-vs-Diff-Aval} presents results obtained by acMOI for an applied field with amplitude of $h_{\text{dc}}^{\text{max}}$ = 2.4~Oe. As revealed in Fig.~\ref{Fig:ACSusc}, such a probe field will lead to the nucleation of flux avalanches in the a-MoSi film for temperatures lower than $T_{\text{th}}$. These abrupt, non-critical-state-like flux penetration events have a characteristic dendritic morphology observable in several MO images in Fig.~\ref{Fig:Reg-vs-Diff-Aval}. Such a strong flux variation leads to the paramagnetic reentrance region in $\chi_{\text{ac}}(T)$ measurements. The acMOI results in the avalanche regime were obtained using a field step of 0.2~Oe and $T$ values of 2.9~K, 3.5~K, 3.8~K, 4.5~K, 5.0~K, 5.5~K, 6.0~K, 6.5~K, and 7.0~K.

The interpretation of Fig.~\ref{Fig:Reg-vs-Diff-Aval} is analogous to that of Fig.~\ref{Fig:Reg-vs-Diff-Smooth}. A series of MO images are taken within the same ac-emulating field cycle both as $T$ is increased after ZFC to 2.9~K and as $T$ is decreased after the magnetic history of the sample is erased above $T_c$. Figure~\ref{Fig:Reg-vs-Diff-Aval}(a) shows images taken at 2.9~K and, hence, below $T_{\text{th}}$. In the first row, the first image reveals that a positive flux avalanche was triggered in the film, advancing further into the interior of the sample than the shallow critical-state-like bright flux front. Then, as revealed by the third image, a new, negative flux avalanche, or anti-avalanche, was triggered, reusing the flux channel created by the first positive avalanche~\cite{Motta2011,Jing2020}. The differential MOI analysis on the third row of Fig.~\ref{Fig:Reg-vs-Diff-Aval}(a) allows us to conclude that the avalanches appearing in the first row were not triggered on the depicted images, but as $h_{\text{dc}}$ was ramped from zero to 2.4~Oe and, then, from 2.4~Oe to $-2.4$~Oe. This is the case because the differential flux distributions do not show any signs of abrupt flux intake by the sample. On the other hand, analysis of the second and fourth rows of Fig.~\ref{Fig:Reg-vs-Diff-Aval}(a) reveals that a positive flux avalanche was triggered in the sample at 2.4~Oe. The differential image allows us to clearly distinguish this specific penetration event from the complex flux landscape presented by the sample.

In the case depicted in Fig.~\ref{Fig:Reg-vs-Diff-Aval}(b), $T$ = 5.5~K $> T_{\text{th}}$. Even though previously triggered avalanches are visible in the first row of images, all flux penetration at this temperature occurs smoothly from the edges of the film. Hence, the very different flux landscapes in the first and second rows lead to similar differential flux patterns, shown in the third and fourth rows of Fig.~\ref{Fig:Reg-vs-Diff-Aval}(b). These results are compatible with those in Fig.~\ref{Fig:ACSusc}, as the sample is in the smooth regime above $T_{\text{th}}$. However, if we now compare the differential images below $T_{\text{th}}$ at 2.9~K, we see that, contrary to the smooth penetration regime, there is not a match between each corresponding image due to the nucleation of flux avalanches. In the Supplemental Material, a video highlights this fact, showing distinct differential flux distributions each time an avalanche occurs in the film either as $T$ is increased or decreased and at different temperatures.

Such a difference in behavior in the avalanche regime, coupled with the unpredictable nature of these events, seems to indicate that there should not be a match between independent $\chi_{\text{ac}}(T)$ measurements. Nevertheless, we do observe in Fig.~\ref{Fig:ACSusc}(b) very similar behaviors as $T$ varies up or down. To understand why this happens, we will rely on the potential of MOI as a quantitative analysis tool, as its spatial resolution allows the study of individual avalanches in a manner that is not possible with standard magnetometers like the MPMS.

\begin{figure}[htbp]
\centering
\includegraphics[width=\columnwidth,keepaspectratio]{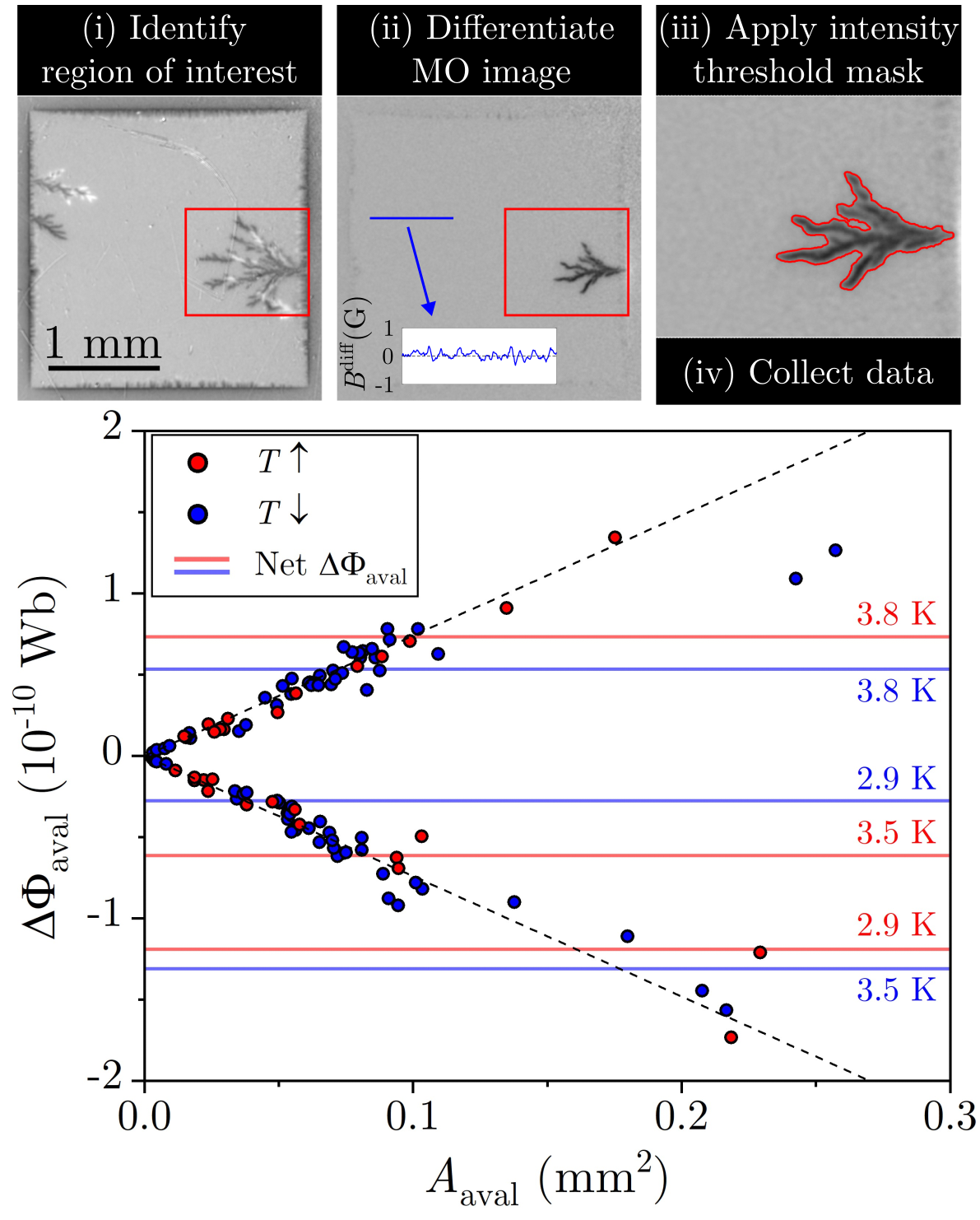}
\caption{Top row: demonstration of the process used to obtain quantitative data on single avalanches. See the main text for a detailed explanation. Main panel: the total magnetic flux of each individual avalanche triggered in the a-MoSi film as the temperature is increased ($T \uparrow$) after ZFC to the base temperature and decreased ($T \downarrow$) from above $T_c$. Results are plotted against the area of the respective avalanches. Solid horizontal lines represent the net $\Delta\Phi_{\text{aval}}$ obtained by summing the flux of all avalanches triggered at the same temperature as $T$ is increased (red) or as $T$ is decreased (blue). Dashed lines are guides to the eye.}
\label{Fig:AvalFlux-vs-Area}
\end{figure}

The first image of the forth row of Fig.~\ref{Fig:Reg-vs-Diff-Aval}(a) indicates how we may extract information on individual avalanches. By differentiating an image in which an avalanche event occurs, we are able to highlight it from the rest of the sample's flux landscape. To harness this possibility and study all avalanches triggered in the a-MoSi film during our measurements, we implemented an algorithm in MATLAB, schematically represented in the top row of Fig.~\ref{Fig:AvalFlux-vs-Area}. As multiple avalanches may occur simultaneously in different parts of the sample, we first select the region of the film which will be analyzed. Then, we differentiate the MO images, resulting in a $B^{\text{diff}}(x,y)$ distribution around zero outside of the smooth flux front and the avalanches. However, avalanches result in much more intense flux variation than critical-state-like penetration. This allows the algorithm to identify every image in which an avalanche was triggered. Moreover, it is possible to clearly separate the avalanches from the rest of the image by applying an intensity threshold mask over the selected avalanche region. This mask can be used either on the directly obtained MO image or on the differential flux distribution, allowing the investigation of different quantities.

Using this algorithm, we analysed all 3474 MO images obtained using the ac-emulated field cycles with $h_{\text{dc}}^{\text{max}}$ = 2.4~Oe. In those, a total of 105 flux avalanches were triggered. Table~\ref{Table:aval-numb} specifies the distribution of these avalanches between different temperatures. It also highlights if the avalanches were triggered while increasing or decreasing $T$ as well as if they are comprised of positive or negative flux. These statistics reveal that many more avalanches occur when the temperature is being reduced from above $T_c$. This difference is related to the established flux landscape within the sample, clearly visible on the second image row in Fig.~\ref{Fig:Reg-vs-Diff-Aval}. As the sample is fully penetrated by vortices, the probability of triggering thermomagnetic instabilities increases~\cite{Qviller2010,Pinheiro2022}. 

\begin{table}[htbp]
\caption{\label{Table:aval-numb} Number of flux avalanches observed in the MO images obtained during the $T \uparrow$ and $T \downarrow$ experiments for field cycles with $h_{\text{dc}}^{\text{max}}$ = 2.4~Oe at different temperatures.}
\begin{ruledtabular}
\begin{tabular}{lllll}
                      & 2.9~K & 3.5~K & 3.8~K & 4.5~K \\
$T \uparrow$ | Positive flux   & 8   & 3   & 3   & 0   \\
$T \uparrow$ | Negative flux   & 8   & 6   & 1   & 0   \\
$T \downarrow$ | Positive flux & 19  & 9   & 10  & 1   \\
$T \downarrow$ | Negative flux & 21  & 10  & 6   & 0  
\end{tabular}
\end{ruledtabular}
\end{table}

Once all avalanches were identified, we may calculate the magnetic flux difference in the sample due to each avalanche, $\Delta\Phi_{\text{aval}}$, by numerically integrating $B^{\text{diff}}(x,y)$. Figure~\ref{Fig:AvalFlux-vs-Area} shows $\Delta\Phi_{\text{aval}}$ as $T$ is increased and decreased as a function of the avalanche area, $A_{\text{aval}}$. Noticeably, the data reveals a temperature-independent linear relationship between $\Delta\Phi_{\text{aval}}$ and $A_{\text{aval}}$, as highlighted by the sloped guides to the eye. This may be understood considering the microscopic nature of the mixed state in type-II superconductors, in which quantized flux vortices permeate the sample. The vortex core size is proportional to the coherence length of the material, $\xi$, whereas the intervortex spacing is related to the penetration depth, $\lambda$~\cite{Blatter1994}. In turn, these quantities evolve in temperature as $(1-T/T_c)^{-1/2}$, which implies that they only vary significantly for temperatures close to $T_c$. Therefore, the density of vortices is nearly constant in the temperature range for which the film is in the avalanche regime, leading to the behavior observed in Fig.~\ref{Fig:AvalFlux-vs-Area}. The slope of the linear relationship is roughly equal to 0.75~mT, indicating fields slightly above the ones used to trigger the avalanches. This difference is explained by the higher flux concentration along the edges of the thin film due to demagnetization effects~\cite{Zeldov1994}.

Moreover, the solid horizontal lines in Fig.~\ref{Fig:AvalFlux-vs-Area} represent the net $\Delta\Phi_{\text{aval}}$ calculated by summing $\Delta\Phi_{\text{aval}}$ for all avalanches that occur at a given set temperature when $T$ is increased (red lines) or decreased (blue lines). Although many more avalanches happen during the $T \downarrow$ experiment, the blue lines reveal that the net flux variation they cause in the sample is comparable to that caused by a single avalanche. The same is true during the temperature increase, as shown by the red lines. This fact is associated with the effects of an ac field cycle on the superconducting film. As can be observed in Fig.~\ref{Fig:Reg-vs-Diff-Aval}, there is a tendency for new avalanches to reuse the flux channel created by previously nucleated avalanches of opposite polarity. Note that bright and dark contrast avalanches are superimposed in the MO images. This same trend has been previously reported both experimentally~\cite{Motta2011} and numerically~\cite{Jing2020}. Such behavior is explained by the attractive nature of the interaction between vortices and antivortices, as well as by the fact that the existing avalanche creates an easy channel of locally reduced critical current density inside the film, facilitating the propagation of magnetic flux. A Supplemental Material video demonstrates that most new avalanches reuse previously existing flux channels. Therefore, these dynamics tend to balance out positive and negative flux variations arising from abrupt penetration events. As the ac susceptibility is measured by averaging the flux variation captured throughout several ac field cycles, the avalanche contributions become very similar in both directions of temperature variation, resulting in the remarkably similar $\chi_{\text{ac}}(T)$ measurements as $T$ is increased and decreased, as shown in Fig.~\ref{Fig:ACSusc}(b).

\section{Quantitative ac susceptibility analysis from MOI}
\label{Sc:qMOI}

In Section~\ref{Sc:acSusc-MOI}, we qualitatively discussed the link between differential MO images and ac susceptibility measurements conducted in the MPMS. In this Section, we demonstrate how MOI can be further utilized as a tool for quantitatively studying ac field-induced effects on superconducting films. The in-phase and out-of-phase components of $\chi_{\text{ac}}$ are obtained by acMOI as a function of $T$, which can then be compared to MPMS measurements.

\begin{figure}[htbp]
\centering
\includegraphics[width=\columnwidth,keepaspectratio]{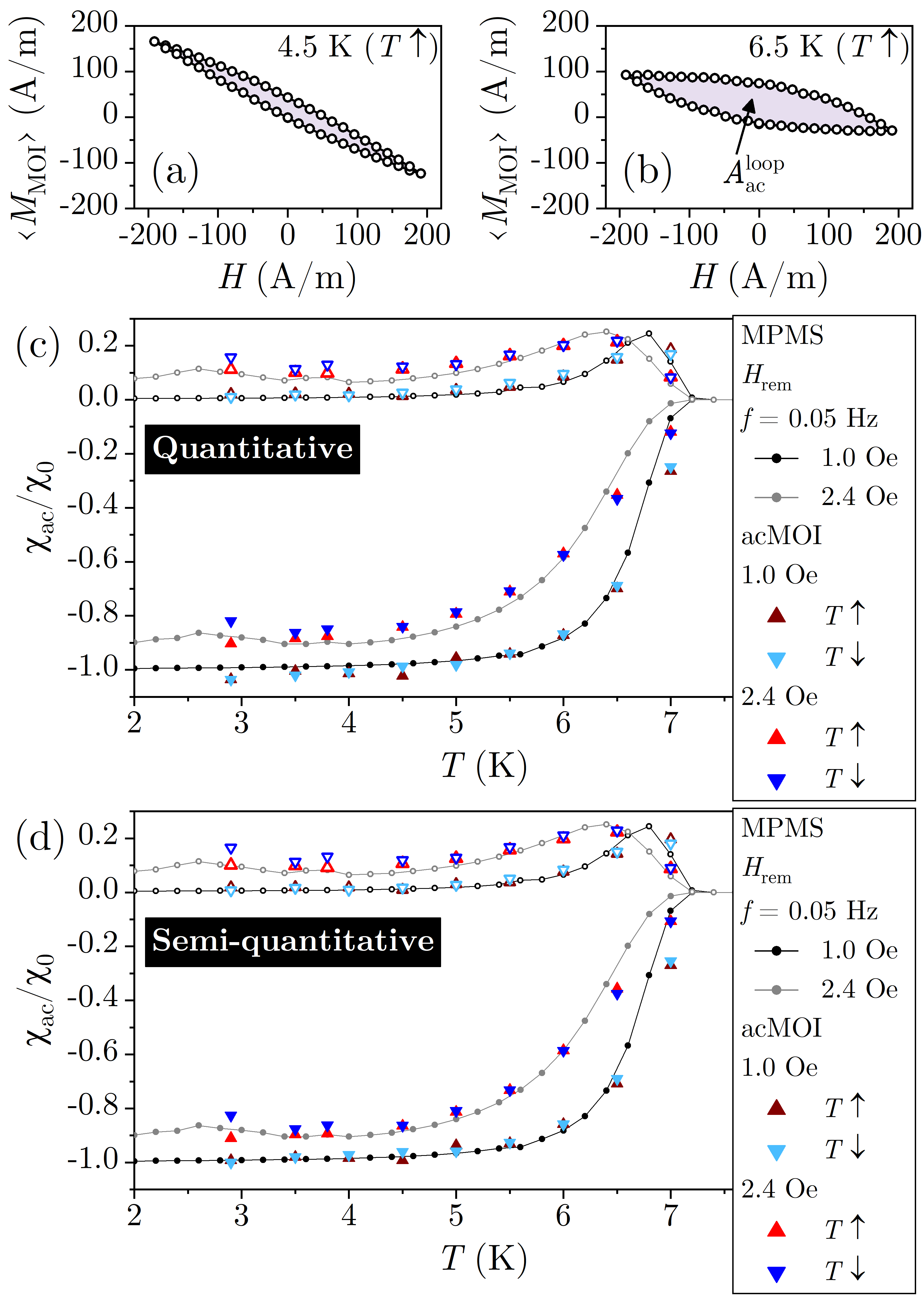}
\caption{Example of typical $M_{\text{MOI}}(h_{\text{dc}})$ loops obtained for measurements as the temperature is increased ($T \uparrow$) at (a) 4.5~K and (b) 6.5~K for $h_{\text{dc}}^{\text{max}}$ = 2.4~Oe $\approx$ 191~A/m. Comparison between the temperature-dependent ac susceptibility of the a-MoSi film under $H_{\text{rem}}$ obtained using the MPMS and (c) a quantitative acMOI analysis and (d) a semi-quantitative acMOI analysis. In both panels, the MPMS measurements are carried at $f$ = 0.05~Hz for better correspondence with the slower MO measurements. MO images are taken as the temperature is increased ($T \uparrow$) and decreased ($T \downarrow$), with ac-emulated field amplitudes of 1.0~Oe and 2.4~Oe, thus in the smooth and avalanche regimes, respectively.}
\label{Fig:Comp-MPMS-MOI}
\end{figure}

To achieve that, let us first be reminded that $\chi_{\text{ac}}'$ is associated with the superconductor inductive response to shield magnetic flux from its interior. Therefore, $\chi_{\text{ac}}'$ captures the evolution of the sample magnetization with an applied magnetic field. On the other hand, $\chi_{\text{ac}}''$ is associated with a resistive response arising from energy losses, caused by the dissipative flux motion within the superconductor. As discussed in Ref.~\cite{Clem1991}, this energy can be gauged by evaluating the area of the $M(h)$ loop, $A_{\text{ac}}^{\text{loop}}$, defined by the application of one ac field cycle. This way, we may obtain the $\chi_{\text{ac}}(T)$ components from ac-emulating MOI cycles as~\cite{Motta2011}
\begin{equation}
    \chi_{\text{ac}}' = \left\langle \frac{\partial \langle M_{\text{MOI}}\rangle}{\partial h_{\text{dc}}} \right\rangle \hspace{12pt} \text{and} \hspace{12pt} \chi_{\text{ac}}'' = \frac{A_{\text{ac}}^{\text{loop}}}{\pi (h_{\text{dc}}^{\text{max}})^2} ,
\label{Eq:acSusc-MOI}
\end{equation}
where the mean magnetization $\langle M_{\text{MOI}} \rangle$ is obtained from the out-of-plane flux density distribution within the sample on a MO image as~\cite{Motta2011}
\begin{equation}
    \langle M_{\text{MOI}}\rangle = \frac{1}{N_{\text{px}}}\sum_{n=1}^{N_{\text{px}}}\left\{ B_n(x,y)/\mu_0 - h_{\text{dc}}\right\} ,
\end{equation}
where $N_{\text{px}}$ is the number of pixels which correspond to the sample within the MO image. These quantities are calculated for each ac-emulating field cycle at a given temperature in SI units as exemplified in Fig.~\ref{Fig:Comp-MPMS-MOI}(a-b), which shows typical $\langle M_{\text{MOI}}\rangle (h_{\text{dc}})$ loops. The results are then averaged for the four cycles to obtain the $\chi_{\text{ac}}(T)$ evolution for the sample.

Figure~\ref{Fig:Comp-MPMS-MOI}(c-d) displays MPMS measurements of $\chi_{\text{ac}}(T)$ for the a-MoSi film using probe fields with $f$ = 0.05~Hz and $h$ = 1.0~Oe and 2.4~Oe, hence in the smooth and avalanche regimes, respectively. Although the $\chi_{\text{ac}}$ analysis is frequency-independent, this $f$ value is chosen to match an ``effective" frequency estimated considering a light exposure time of 200~ms during the acMOI measurements, which, in turn, is used to optimize the image contrast. The MPMS data in Fig.~\ref{Fig:Comp-MPMS-MOI} is averaged from eight successive field cycles [see Eqs.~\eqref{Eq:MPMS-acFit} and \eqref{Eq:MPMS-m'm''}]. Although this differs from the four cycles used in the acMOI measurements, Appendix~\ref{App:numcycles} demonstrates that MPMS results are equivalent for measurements conducted with these numbers of field cycles. Thus, in Fig.~\ref{Fig:Comp-MPMS-MOI}(c), the MPMS measurements are compared to $\chi_{\text{ac}}'$ and $\chi_{\text{ac}}''$ quantitatively obtained from the acMOI measurements using Eq.~\eqref{Eq:acSusc-MOI}, both as the temperature is increased after ZFC and as $T$ is decreased from above $T_c$. There are two main observations in Fig.~\ref{Fig:Comp-MPMS-MOI}(c). The first is that, despite limitations in the measurement resolution in comparison to SQUID magnetometers and the presence of defects on the MO indicator which could compromise the result, acMOI captures with high fidelity the behavior of both components of $\chi_{\text{ac}}(T)$, specially at the lower temperatures and ac field amplitudes. When $T$ approaches $T_c$, however, the lower contrast of the MO images induce larger errors, therefore, the acMOI data points at 7~K are detached from those obtained using the MPMS. The second observation is that acMOI captures exceptionally well the independence of  $\chi_{\text{ac}}(T)$ on thermomagnetic history in the smooth regime, as $\chi_{\text{ac}}'$ and $\chi_{\text{ac}}''$ are mostly superimposed in Fig.~\ref{Fig:Comp-MPMS-MOI}(c). When the film is in the avalanche regime, acMOI also captures the paramagnetic reentrance observed in the MPMS measurements. However, it appears that the technique is more susceptible to differences in the flux landscape in the sample, as measurements conducted as the temperature was decreased resulted in slightly lower values of $|\chi_{\text{ac}}'|$. Nonetheless, within the natural limitations of the technique, it is possible to accurately investigate the ac susceptibility of a superconducting thin film using ac-emulating MOI.

A semi-quantitative approach can also be used to obtain $\chi_{\text{ac}}(T)$ from acMOI. As highlighted by Eq.~\eqref{Eq:acSusc-MOI}, the sample magnetization is the crucial ingredient in the calculation of $\chi_{\text{ac}}'$ and $\chi_{\text{ac}}''$. $M$, however, is a global parameter, describing the average behavior of the sample. In Fig.~\ref{Fig:Comp-MPMS-MOI}(c), we obtained this quantity from the local flux density distribution in the film. If we remember that raw MOI data is an intensity count, we may define a mean intensity for each MO image, $\langle I(x,y) \rangle$. Then, using measurements performed above $T_c$, such that the sample magnetization does not interfere with the flux distribution, we may find a relationship between an applied magnetic field $H$ and $\langle I(x,y) \rangle$. Considering that, above $T_c$, $M$ = 0 and $H = B/\mu_0$, the mean flux density distribution $\langle B \rangle$ can be found by fitting an empirical polynomial relationship between $\langle B \rangle$ and $\langle I(x,y) \rangle$~\footnote{For the data presented in this work, we found by inspection that a third degree polynomial successfully described the behavior of the mean intensity with the applied magnetic field.}. The influence of defects on the MO indicator can be minimized by subtracting the zero-field background from all images.

Once the images are calibrated, the mean sample magnetization in each MO image within an ac-emulating field cycle can be calculated as
\begin{equation}
    \langle M \rangle = \langle B \rangle/\mu_0 - h_{\text{dc}} .
\end{equation}
Using $\langle M \rangle$ and Eq.~\eqref{Eq:acSusc-MOI}, we obtained the $\chi_{\text{ac}}(T)$ results shown in Fig.~\ref{Fig:Comp-MPMS-MOI}(d). The results are completely analogous and very similar to those depicted in Fig.~\ref{Fig:Comp-MPMS-MOI}(c), demonstrating the robustness of MOI as a tool to gauge $\chi_{\text{ac}}(T)$.

\section{Erasing flux avalanches}
\label{Sc:ErasingAval}

\begin{figure}[htbp]
\centering
\includegraphics[width=\columnwidth,keepaspectratio]{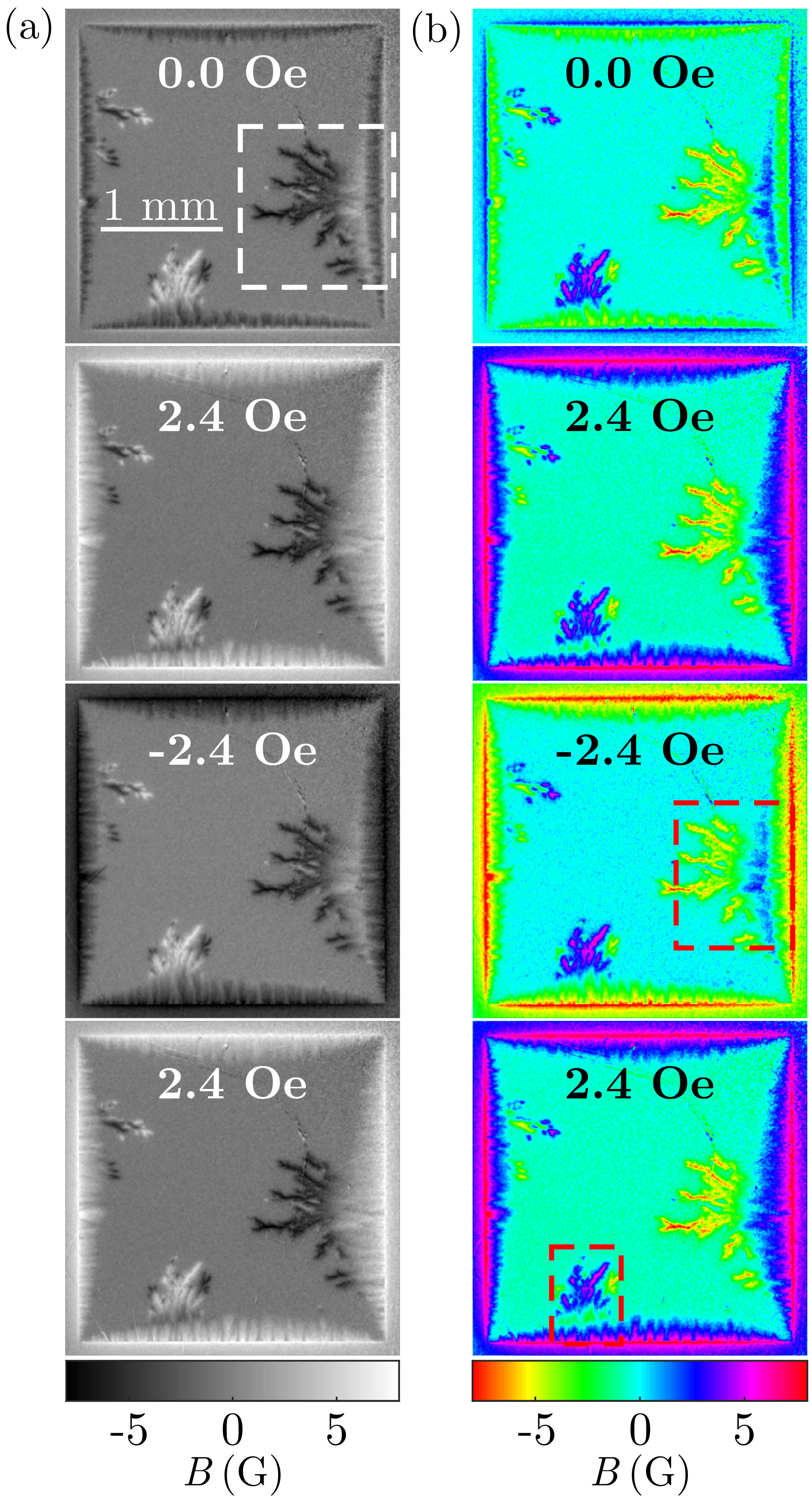}
\caption{MOI of a-MoSi film at 5~K as the temperature is increased after ZFC. The images were captured during ac-emulating field cycles with $h_{\text{dc}}^{\text{max}}$ = 2.4~Oe and demonstrate how an incoming flux front interacts with previously established avalanches. Panels (a) and (b) show the same MO images with different color scales to highlight different features of the flux penetration dynamics. Dashed rectangles highlight regions in which it is possible to observe the vortex-antivortex annihilation zone.}
\label{Fig:ErasingAval}
\end{figure}

Let us now discuss a side benefit of using quantitative MO data to gain insight into the interaction between an incoming magnetic flux front and the region where an avalanche previously took place. In Fig.~\ref{Fig:Reg-vs-Diff-Aval}(b), arrows indicate regions in differential images taken at 2.4~Oe and $-$2.4~Oe in which positive and negative flux, respectively, penetrate further into the sample than elsewhere. Figure~\ref{Fig:ErasingAval} sheds light on this dynamics by highlighting results obtained for the a-MoSi sample at 5~K as $T$ is increased from the base temperature after ZFC. Panels (a) and (b) show the same MO images side-by-side, only with different color scales. This is done to evidence different aspects of the flux penetration dynamics.

The first image of Fig.~\ref{Fig:ErasingAval}, taken at 0~Oe, is captured before the ac-emulating magnetic field is applied to the film at 5~K. Therefore, the depicted flux landscape is a result of the 16 ac-emulating field cycles applied to the film at the four previous temperature steps. Noticeably, a number of flux avalanches took place, resulting in the characteristic dendritic flux-filled regions observed in the sample. On the second row, $h_{\text{dc}}$ is increased to 2.4~Oe for the first time at 5~K. As previously discussed, this will result in the penetration of a positive, smooth flux front from the edges toward the center of the film. Figure~\ref{Fig:ErasingAval}(a) illustrates an interesting characteristic of the dynamics of how this flux front interacts with avalanches previously triggered in the film. First, notice the presence of a large negative flux avalanche on the right edge of the film framed by the dashed white rectangle. Then, we may observe that the positive flux front penetrates deeper into the sample where it interacts with the negative avalanche than elsewhere\textemdash compare, for instance, the penetration from the right edge with that from the top edge of the sample. Additionally, a medium-sized positive flux avalanche had previously occurred on the bottom-left edge of the film. Near that avalanche, the positive flux front has a shallower penetration than on the right side of the bottom edge. The explanation for such a difference in the flux penetration lies in the nature of the attractive interaction between vortices and antivortices~\cite{Chaves2021}, leading to the deeper penetration of the flux front coming from the right edge on the second image of Fig.~\ref{Fig:ErasingAval}(a). However, if the incoming flux has the same polarity as the pinned flux, it will be repelled, causing the shallower penetration of the positive flux front over the positive avalanche on the bottom edge of the sample.

Moreover, vortices and antivortices will be annihilated if they come in close contact, leaving behind a region of zero net magnetic flux on the superconductor. Figure~\ref{Fig:ErasingAval}(b) allows us to visualize such vortex-antivortex annihilation regions. Using again the second image of the depicted sequence as a reference, we may look at the right edge of the sample, where the positive (dark-blue) flux front penetrates over the negative (yellow) avalanche. Then, we notice the presence of a zero flux (light-blue) region between the flux front and the avalanche. As vortices penetrate the sample from the right edge, they encounter previously pinned antivortices, leading to mutual annihilation. The resulting flux-free region is then filled by new incoming vortices which, in turn, will be annihilated with further pinned antivortices, in a process that enables the positive flux penetration as the field is increased up to $h_{\text{dc}}^{\text{max}}$.

In the next step of the ac-emulating field cycle, the field is reduced to $-h_{\text{dc}}^{\text{max}}$. Then, negative (yellow) flux will penetrate the sample from the edges. As observed in the third row of Fig.~\ref{Fig:ErasingAval}, negative flux penetrates less from the right edge of the sample than the positive flux front did. Moreover, we observe that the negative flux further penetrates over the positive flux avalanche that previously occurred at the bottom edge of the sample. Thus, the negative flux penetration dynamics follow the same behavior observed when a positive flux front penetrates the film. Accordingly, the third image of Fig.~\ref{Fig:ErasingAval}(b) reveals a vortex-antivortex annihilation zone between the incoming negative flux front and the deeper positive front. Then, inside the dashed red rectangle, we observe beginning from the edge of the sample: a negative flux region, a first annihilation zone, a positive flux region, a second annihilation zone, and, finally, the negative deeply pinned flux where the avalanche propagated through the film. In the fourth image of the depicted sequence, the applied field is once again increased to $h_{\text{dc}}^{\text{max}}$, leading to positive flux penetration. Now, along the bottom edge, the incoming positive flux penetrates less than the established negative flux over the positive avalanche. This creates the region highlighted by the dashed rectangle on the fourth image of Fig.~\ref{Fig:ErasingAval}(b), where it is possible to observe a positive flux region, an annihilation zone, a negative flux region, another annihilation zone, and the positive flux pinned after the avalanche penetrated deep into the sample. The Supplemental Material presents a video highlighting the interaction between an incoming flux front with the pre-established avalanches in Fig.~\ref{Fig:ErasingAval} at different moments of the ac-emulating field cycle.

\begin{figure}[htbp]
\centering
\includegraphics[width=\columnwidth,keepaspectratio]{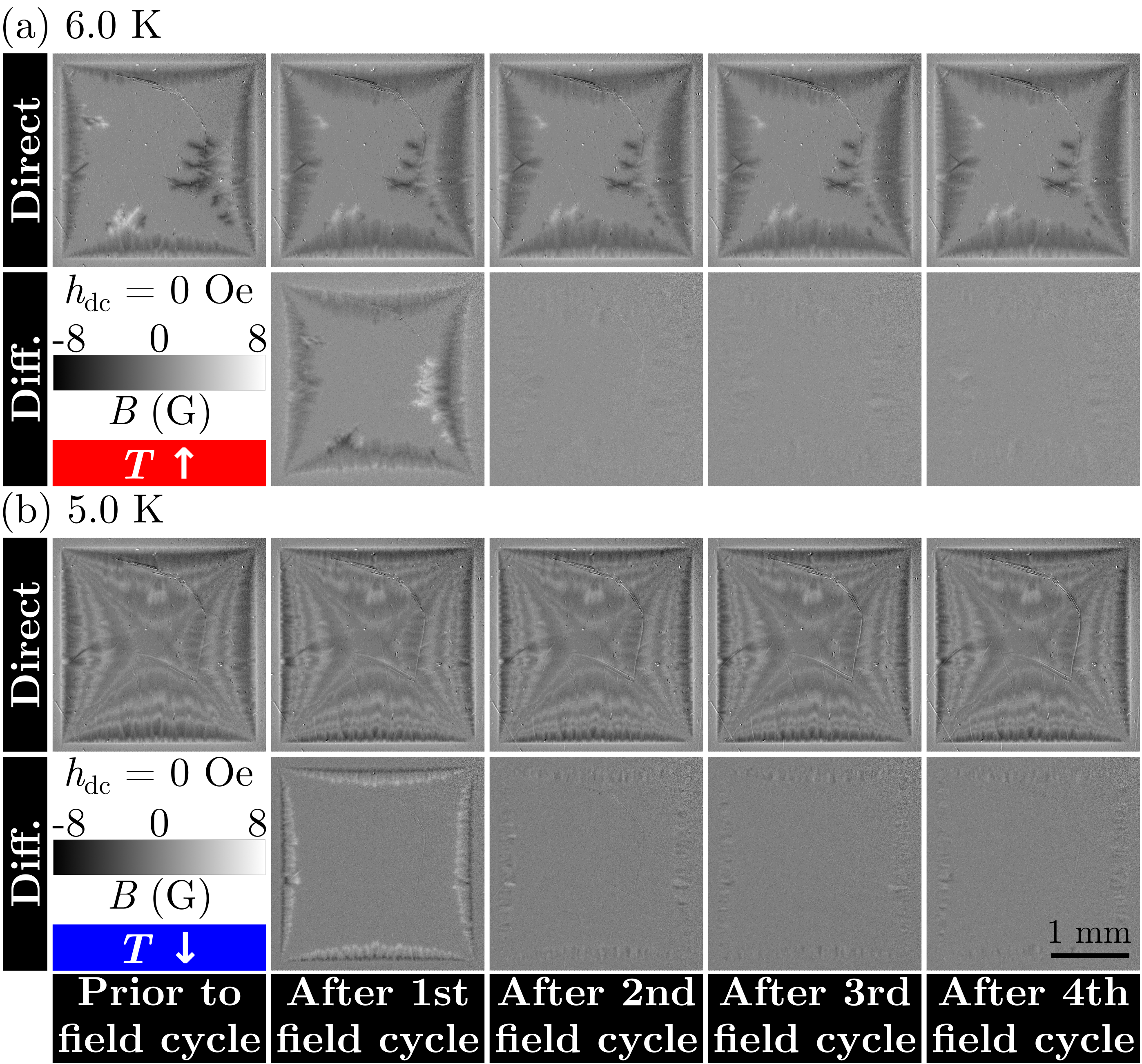}
\caption{MOI of a-MoSi film at (a) 6~K as $T$ is increased after ZFC and (b) 5~K as $T$ is decreased from above $T_c$. The images were captured after ac-emulating field cycles with $h_{\text{dc}}^{\text{max}}$ = 2.4~Oe were completed, i.e., under $h_{\text{dc}}$ = 0~Oe. The first row shows direct measurements whereas the second row shows differential results. The differential images were obtained by subtracting the images after a field cycle from those obtained after the previous field cycle, i.e., the one in the previous column on the first row.}
\label{Fig:AfterCycle}
\end{figure}

The flux penetration dynamics in the smooth penetration regime revealed by Fig.~\ref{Fig:ErasingAval} hints at a different aspect of ac susceptibility measurements. To wit, Fig.~\ref{Fig:ErasingAval}(a) shows two panels with $h_{\text{dc}}$ = 2.4~Oe where no clear differences are observed in the penetrated flux landscape. Figure~\ref{Fig:AfterCycle} further explores this aspect of the results both as the temperature is increased after ZFC [Fig.~\ref{Fig:AfterCycle}(a), at 6~K] and as it is decreased from above $T_c$ [Fig.~\ref{Fig:AfterCycle}(b), at 5~K]. The first row of both panels shows MO images obtained before the ac-emulating magnetic field is applied at the indicated temperature followed by the flux landscape captured at the end of each field cycle, when the a-MoSi is under $h_{\text{dc}}$ = 0~Oe. The second row shows differential MO images obtained by subtracting the flux landscape after field cycle $N$ by that after cycle $N-1$. As established, flux penetration differs when the field is applied at different temperatures. Thus, a different flux pattern is revealed after the first cycle when compared to the previously pinned landscape, as evidenced by the first differential image in both panels. However, as the applied field reaches $h_{\text{dc}}^{\text{max}}$ (or $-h_{\text{dc}}^{\text{max}}$), the penetrated positive (or negative) flux front reaches its maximum depth into the sample for those specific measurement conditions. Therefore, there is no sensitive difference between the flux landscapes observed at equivalent $h_{\text{dc}}$ at the subsequent field cycles after the field reaches its maximum value. This is evidenced by the last three differential images in both panels. In the Supplemental Material, an accompanying video shows that these dynamics are observed in all images captured within the four field cycles for the measurements presented in Fig.~\ref{Fig:AfterCycle}(a). Therefore, in the smooth regime, the important dynamic aspects of flux penetration into superconducting samples are restricted to occur during the first field cycle. This naturally explains the observed independence of $\chi_{\text{ac}}$ on the number of averaged field cycles in MPMS measurements, as reported in Appendix~\ref{App:numcycles}.

\section{Conclusions}
\label{Sc:conclusions}

We have investigated the ac magnetic susceptibility of a superconducting thin film with lateral dimensions in the millimeter range. Standard global ac magnetometry measurements of frequency-independent first harmonic $\chi_{\text{ac}}(T)$ reveal that the sample exhibits a paramagnetic reentrance related to the abrupt magnetic flux intake experienced during a flux avalanche event. Despite the stochastic nature of these avalanches, their effect on $\chi_{\text{ac}}(T)$ is nearly insensitive to the sample thermomagnetic history. We employ quantitative ac-emulating magneto-optical imaging to uncover the reasons behind this fact. In the smooth penetration regime, the indistinguishability of $\chi_{\text{ac}}(T)$ measured as the temperature is increased from 2~K or decreased from above $T_c$ is explained using differential MO images highlighting that the flux variation within the sample during an ac cycle is independent of the previously established flux landscape. The same is not true in the presence of flux avalanches. Nevertheless, we demonstrate that new avalanches preferentially nucleate along previously established and frozen avalanche regions of opposite polarity. By quantifying the flux variation due to each single avalanche, we find out that this process leads to similar contributions as $T$ is increased or decreased. We thus correlate these findings to the similar $\chi_{\text{ac}}(T)$ behavior in the avalanche regime, independently of the thermomagnetic history of the sample. Moreover, we use acMOI to quantitatively gauge $\chi_{\text{ac}}(T)$ in superconductors obtaining excellent agreement with standard global measurements, particularly at low temperatures and probe field amplitudes. Although the results have been obtained for an a-MoSi film, they are of total generality and, in principle, applicable to any kind of type-II superconductor, even those with high critical temperatures. We also take advantage of the technique to locally resolve regions of vortex-antivortex annihilation, explaining how an income flux front interacts with previously nucleated avalanches. This interplay also allows us to visualize that, after the ac field reaches its maximum amplitude in both field polarities, no new features are observed for subsequent field cycles, explaining the observed independence of $\chi_{\text{ac}}$ on the number of cycles averaged to obtain the results. Therefore, by analyzing the history-independent $\chi_{\text{ac}}(T)$ of an a-MoSi sample, we demonstrate that acMOI is an effective technique to quantitatively study frequent-independent ac magnetic field effects in superconducting materials. This was recently employed to explain the impact of flux dynamics and, in particular, avalanches, on the resonance frequency of large-area superconducting coplanar waveguide resonators~\cite{Nulens2023}.

\acknowledgements{This work was partially supported by Coordenação de Aperfeiçoamento de Pessoal de Nível Superior -- Brasil (CAPES) -- Finance Code 001, the São Paulo Research Foundation (FAPESP, Grant No. 2021/08781-8), the National Council for Scientific and Technological Development (CNPq, Grants No. 431974/2018-7 and 316602/2021-3) and by the UK EPSRC through Grant EP/I036303/1. 

D.A.D.C. and J.C.C.F. contributed equally to this work.}

\appendix

\section{Ac susceptibility dependency on the drive frequency}
\label{App:frequency}

\begin{figure}[htbp]
\centering
\includegraphics[width=\columnwidth,keepaspectratio]{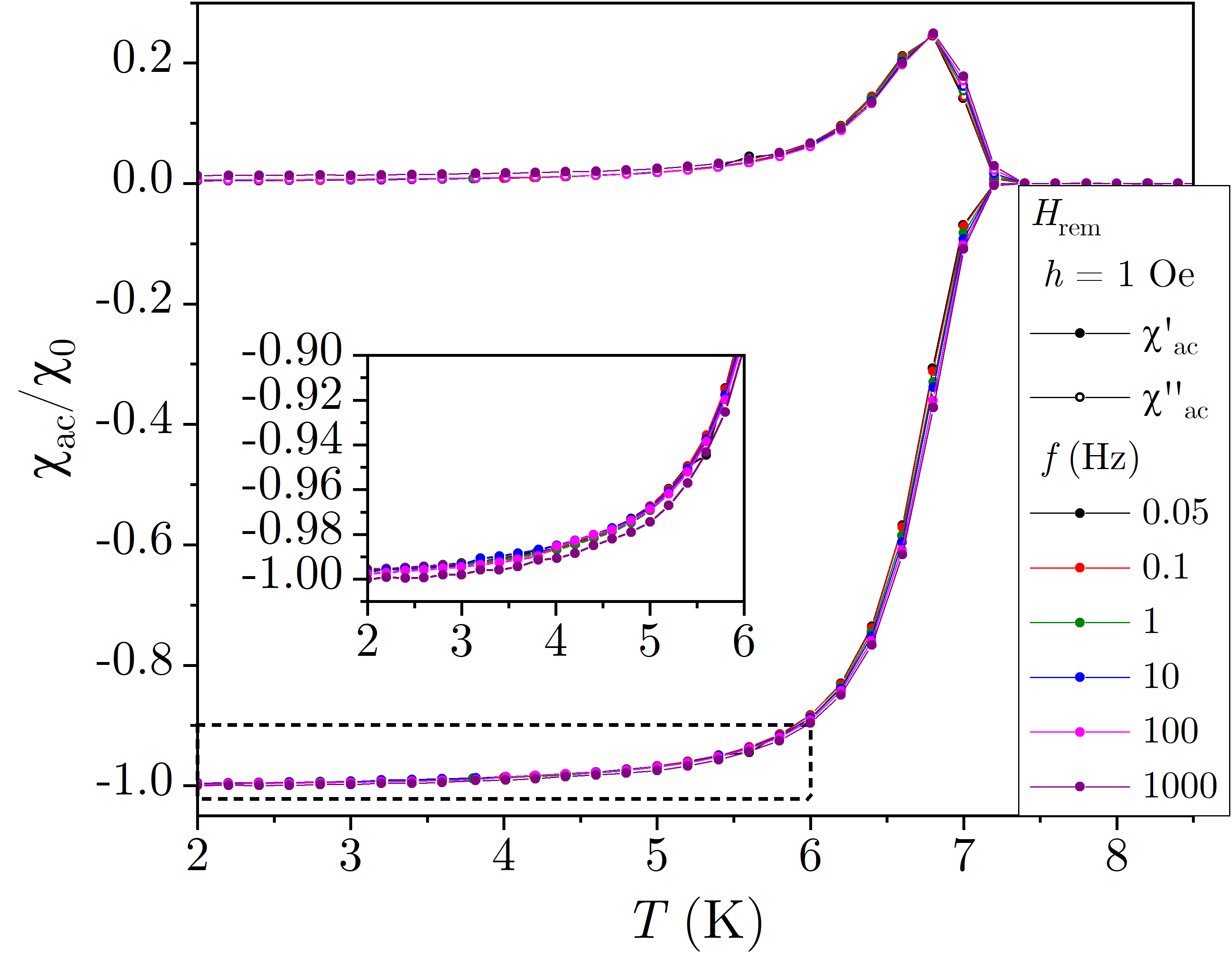}
\caption{Temperature-dependent ac susceptibility under $H_{\text{rem}}$ of a-MoSi film obtained using the MPMS. The probe field amplitude is kept constant at $h$ = 1~Oe while $f$ is varied between 0.05~Hz to 1000~Hz. Inset shows an amplification of the graph region highlighted by the dashed rectangle in the main panel.}
\label{Fig:frequency}
\end{figure}

Figure~\ref{Fig:frequency} shows $\chi_{\text{ac}}(T)$ measurements performed in the MPMS varying the magnetic field drive frequency between 0.05~Hz and 1000~Hz. The results are normalized by $\chi_0$ obtained from the $f$ = 1000~Hz curve. In the investigated frequency range, $\chi_{\text{ac}}(T)$ shows no significant variations for different values of $f$.

\section{Ac susceptibility dependence on the number of field cycles}
\label{App:numcycles}

\begin{figure}[htbp]
\centering
\includegraphics[width=\columnwidth,keepaspectratio]{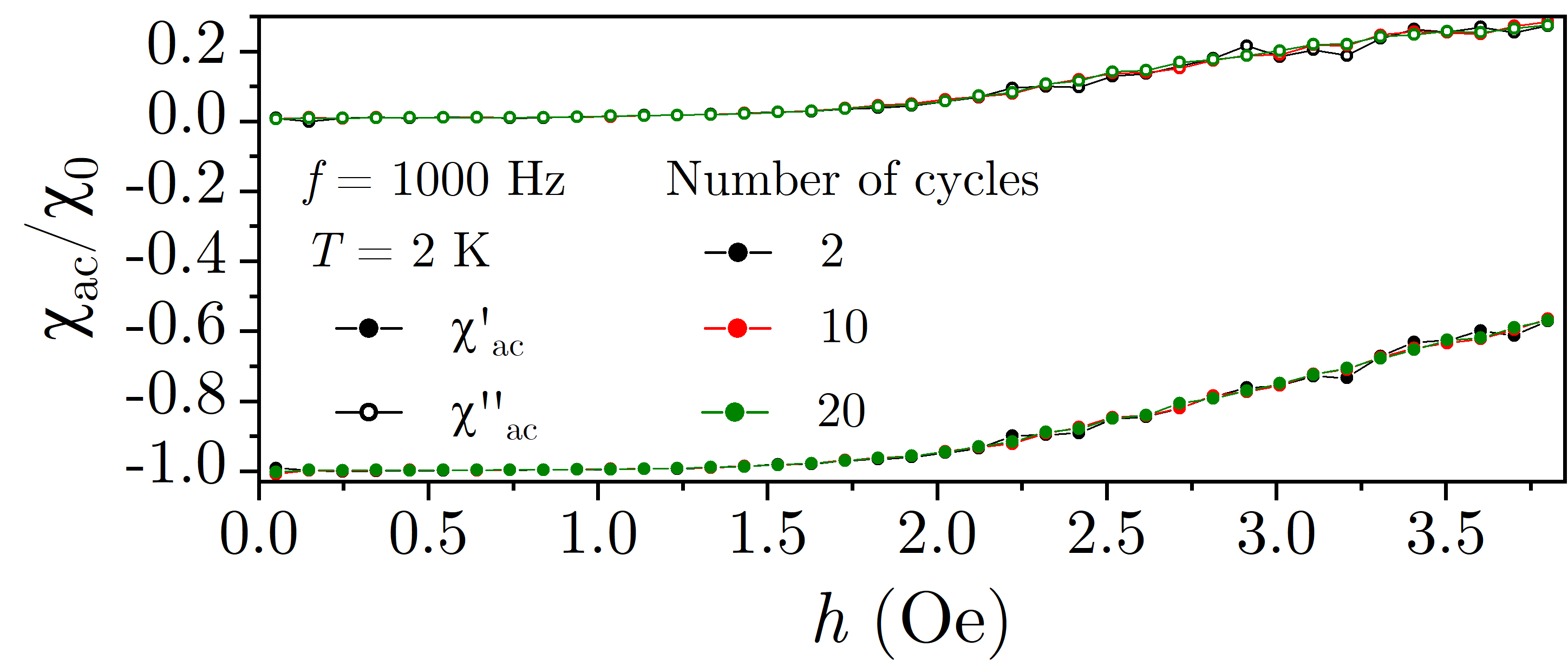}
\caption{Ac susceptibility of a-MoSi film  under $H_{\text{rem}}$ as a function of $h$ obtained using the MPMS. The different measurement runs reflect results obtained by averaging different number of field cycles. All measurements were carried out at $f$ = 1000~Hz and $T$ = 2~K.}
\label{Fig:numcycles}
\end{figure}

Figure~\ref{Fig:numcycles} shows three different measurements of the ac susceptibility of the a-MoSi sample as a function of $h$. The data is obtained varying the number of field cycles used by the MPMS to average the magnetic moment [see Eqs.~\eqref{Eq:MPMS-acFit} and \eqref{Eq:MPMS-m'm''}]. The $h$ range explored depicts the full limit of the MPMS. If the sample is in the smooth penetration regime, i.e., $h <$ 2.0~Oe for $T$ = 2~K, Fig.~\ref{Fig:numcycles} quantitatively shows that $\chi_{\text{ac}}$ is independent of the number of field cycles. In the avalanche regime, small variations are observed due to the stochastic nature of the abrupt flux penetration events.

\newpage

\bibliography{references.bib}

\end{document}